\documentclass[nonacm,sigplan,10pt]{acmart}
\usepackage{amsmath}

\usepackage{amssymb,amsfonts}
\usepackage{algorithm}
\usepackage{algpseudocode}
\usepackage{graphicx}
\usepackage{textcomp}
\usepackage{xcolor}
\usepackage{microtype}
\usepackage{subcaption}
\usepackage{booktabs}
\usepackage{longtable}
\usepackage{makecell}
\usepackage{multirow}
\usepackage{amsmath}
\usepackage{amsfonts}    
\usepackage{mathrsfs}
\usepackage{amsthm}
\usepackage{hyperref}
\usepackage[switch]{lineno}
\usepackage{fancyhdr}
\usepackage{pifont}
\newcommand{\name}{\texttt{REFT}}

\newcommand{\names}{\texttt{REFT-save}}
\newcommand{\namel}{\texttt{REFT-load}}

\newtheorem{assumption}{Assumption}

\begin{document}
\title{Fault-Tolerant Hybrid-Parallel Training at Scale with Reliable and Efficient In-memory Checkpointing}

\author{Yuxin Wang*}
\affiliation{%
  \institution{Hong Kong Baptist University}
  \city{Hong Kong}
  \country{China}
}

\author{Xueze Kang*}
\affiliation{%
  \institution{The Hong Kong University of Science and Technology (Guangzhou)}
  \city{Guangzhou}
  \country{China}
}

\author{Shaohuai Shi}
\affiliation{%
  \institution{Harbin Institute of Technology (Shenzhen)}
  \city{Shenzhen}
  \country{China}
}

\author{Xin He}
\affiliation{%
  \institution{Hong Kong Baptist University}
  \city{Hong Kong}
  \country{China}
}

\author{Zhenheng Tang}
\affiliation{%
  \institution{Hong Kong Baptist University}
  \city{Hong Kong}
  \country{China}
}

\author{Xinglin Pan}
\affiliation{%
  \institution{The Hong Kong University of Science and Technology (Guangzhou)}
  \city{Guangzhou}
  \country{China}
}

\author{Yang Zheng}
\affiliation{%
  \institution{Huawei Technologies Co., Ltd.}
  \city{Shenzhen}
  \country{China}
}

\author{Xiaoyu Wu}
\affiliation{%
  \institution{Huawei Technologies Co., Ltd.}
  \city{Shenzhen}
  \country{China}
}

\author{Amelie Chi Zhou}
\affiliation{%
  \institution{Hong Kong Baptist University}
  \city{Hong Kong}
  \country{China}
}

\author{Bingsheng He}
\affiliation{%
  \institution{National University of Singapore}
  \country{Singapore}
}

\author{Xiaowen Chu}
\thanks{*: Equal contribution; Corresponding author: xwchu@ust.hk}
\affiliation{%
  \institution{The Hong Kong University of Science and Technology (Guangzhou)}
  \city{Guangzhou}
  \country{China}
}

\begin{abstract}



To efficiently scale large model (LM) training, researchers transition from data parallelism (DP) to hybrid parallelism (HP) on GPU clusters, which frequently experience hardware and software failures. Existing works introduce in-memory checkpointing optimizations that snapshot parameters to device memory for rapid failure recovery. However, these methods introduce severe resource competition between checkpointing and training, which can work under DP but can hardly scale under resource-intensive HP.

To ensure low checkpointing overhead for hybrid-parallel training, this paper introduces a distributed in-memory checkpointing system with near-zero in-memory saving overhead. It strives from two aspects to mitigate the on-host resource competition caused by in-memory checkpointing: (1) It introduces Hierarchical Asynchronous Snapshotting Coordination in the checkpoint saving stage. This approach uses three-level asynchronous on-device scheduling to enhance parallelism between snapshotting and training, thereby minimizing snapshotting overhead.
(2) It proposes Hybrid In-memory Checkpoint Protection to enhance checkpoint completeness during hardware failures. Unlike methods that require inter-node communications, which may block training under HP, it creates intra-node redundancy with efficient resource utilization, protecting training against hardware failures with minimal overhead.
With these methods, this work enables fast restart for failed HP training with Distributed In-memory Checkpoint Loading, bypassing inefficiencies in NFS reads. In our evaluation, we achieve zero in-memory checkpoint saving overhead on Frontier while training Llama-2-34B on 256 MI250X devices (512 GPUs).


\end{abstract}
\maketitle
\pagestyle{plain}

\section{Introduction}\label{sec:introduction}
Pretraining of large models (LMs) such as GPT~\cite{brown2020language, radford2018improving}, T5~\cite{raffel2019exploring}, Megatron~\cite{narayanan-sc21}, Llama~\cite{touvron2023llama}, and OPT~\cite{zhang2022opt} requires extensive GPU resources~\cite{narayanan-sc21,tang2023fusionai} and is also error-prone~\cite{zhang2022opt}. 

For instance, the OPT-175B training~\cite{zhang2022opt} employs significant computational resources of 992 80GB A100 GPUs. The training process experienced frequent failures, resulting in over \textbf{105} restarts over \textbf{60} GPU days~\cite{zhang2022opt}. The longest healthy training period was only \textbf{2.8} days. Hardware issues~\cite{234916,jiang2024megascale,tpuv4,295545}, e.g., overheating and power failures, together with software failures~\cite{234916,jiang2024megascale,tpuv4,295545}, e.g., MPI~\cite{Forum_1994} errors and checkpointing errors~\cite{234916}, are the main causes of failures.
Each failure in GPU clusters causes training interruptions and loss of all parameters in volatile GPU memory~\cite{234916}.

Checkpointing is a commonly used method to guard training against failures. To reduce the overhead during checkpoint saving and loading~\cite{narayanan-sc21}, asynchronous~\cite{mohan_checkfreq_nodate,jiang2024megascale,torchsnapshot} and in-memory checkpointing methods~\cite{jiang2024megascale,gemini} have been developed for LM training. The in-memory checkpointing process includes two parts: snapshotting and in-memory protecting. 
Snapshotting captures the current state of the model, including model parameters, optimizer states, and any other relevant information required to resume training, while protecting ensures the state is safely maintained in memory.
Both snapshotting and protecting can interfere with training~\cite{gemini}. Thus, optimizing the two processes to reduce the overhead of in-memory checkpointing is crucial to fault-tolerant LM training.

\begin{figure}[tb]
    \centering
    \begin{subfigure}{0.23\textwidth}
        \centering
        \includegraphics[width=\textwidth]{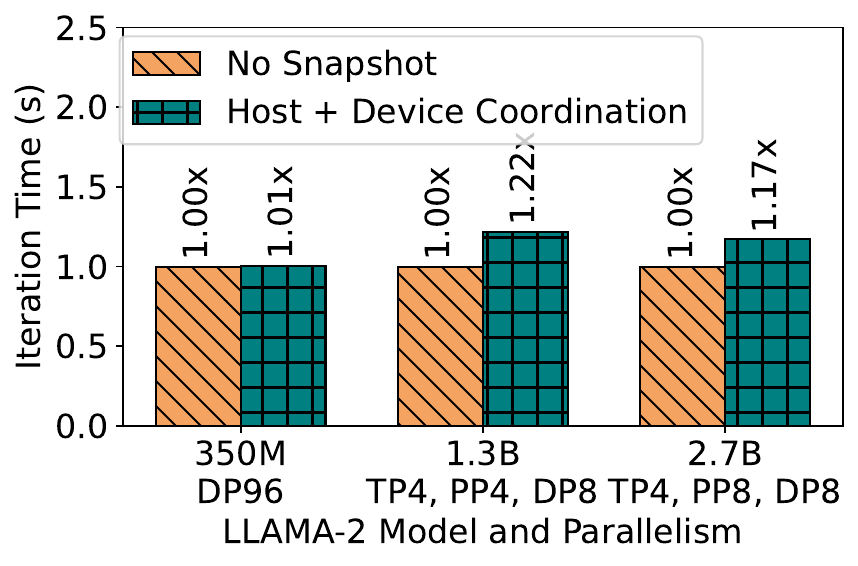}
        \caption{Iteration time with asynchronous snapshotting v.s. without snapshotting}
        \label{fig:async1}
    \end{subfigure}
    \hfill
    \begin{subfigure}{0.23\textwidth}
        \centering
        \includegraphics[width=\textwidth]{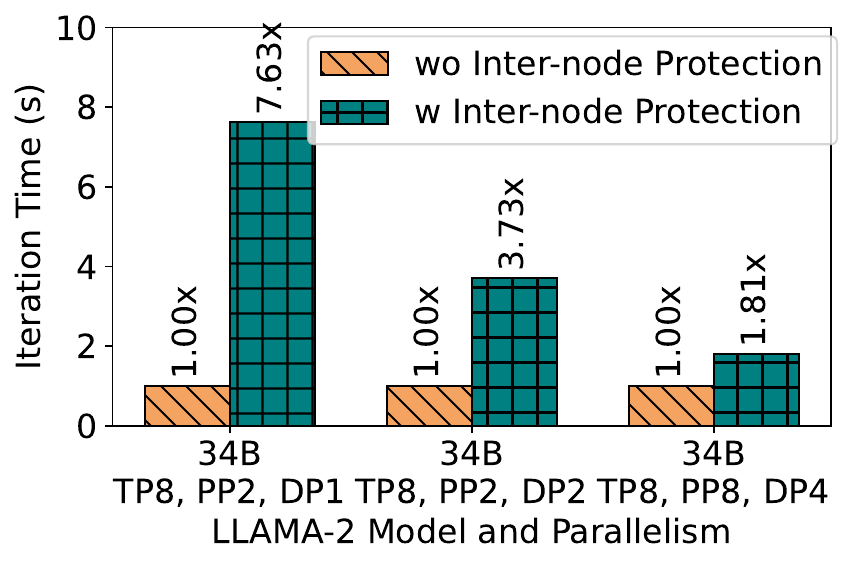}
        \caption{Iteration time with in-memory checkpoint protecting v.s. without protecting}
        \label{fig:sub2}
    \end{subfigure}
    \caption{Existing in-memory checkpoint snapshotting and protecting methods introduce severe overhead to LM training. Details of optimizations will be discussed in Section~\ref{sec:back}. 
    }
    \label{fig:intro:sota}
\end{figure}

Existing asynchronous and in-memory checkpointing methods are primarily designed for data-parallel (DP) training with small models~\cite{mohan_checkfreq_nodate} or ZeRO-3 DP~\cite{rajbhandari2020zero} with limited scalability.
When it comes to large model training under hybrid parallelism (HP), where both model size and system scale increase, existing in-memory checkpointing has shown limitations. 
For example, Figure~\ref{fig:intro:sota} shows the iteration time of training Llama-2 models with different numbers of parameters and using different parallelism. As shown in Figure~\ref{fig:intro:sota}(a), when training with 96 V100 GPUs, existing asynchronous snapshotting~\cite{jiang2024megascale,gemini} introduces low overhead (i.e., 1\%) in DP-only setting, while the overhead increases to 22\% when scaling up the model size and training with HP. More critically, Figure~\ref{fig:intro:sota}(b) reveals the overhead of existing in-memory checkpoint protecting methods~\cite{gemini} when training synthetic Llama-2 models on 8 -- 32 MI250X GPUs. Results have shown that the inter-node communication time consumed during checkpoint protection can be even higher than that of training time (up to 6.63$\times$ of iteration time).
This motivates us to design new snapshotting and protecting optimizations to reduce in-memory checkpointing overhead and ensure fast LM training and failure recovery.

To address the above issues, this paper proposes~\name, an in-memory checkpointing system that reliably and efficiently utilizes volatile host memory to protect snapshots for fast recovery. \name~contains two main components, including \names~for in-memory checkpoint saving and \namel~for in-memory checkpoint loading. \name~exhibits the following capabilities to reduce the checkpointing overhead under the resource-intensive HP:
\begin{itemize}
    \item Efficient Snapshotting: We introduce \textit{Hierarchical Asynchronous Snapshotting (HAS)} that effectively utilizes device idle time while reducing resource competition in hybrid parallel training. This allows large-scale training while snapshotting with near-zero overhead.
    \item Efficient and Reliable In-Memory protecting: \name~minimizes the redundant protecting overhead, automatically adjusting the schemes to generate different volumes of redundancy under different hybrid parallel settings. 
    It introduces Asynchronous Redundant Copying (ARC), Asynchronous Erasure Coding (AEC),  and Asynchronous Optimizer Recomputing (AOR) to enhance the completeness of distributed checkpoints. These methods are powered by HAS during redundant snapshotting, adding little to no additional overhead.
\end{itemize}

We build \name~on PyTorch~\cite{pytorch} and DeepSpeed~\cite{deepspeed-sc20} and evaluate it using Llama-2~\cite{touvron2023llama} models on both NVIDIA Cluster and Frontier to 
demonstrate its general applicability for different GPU hardware. During pretraining Llama-2-6.7B on
512 V100-SXM2-16GB GPUs, REFT-save achieves 5\% to zero overhead compared to the state-of-the-art in-memory protection method. It also achieves zero in-memory checkpoint saving overhead on
Frontier while training Llama-2-34B on 256 MI250X devices, 512 GPUs. REFT has been integrated into an industrial LM training platform, which has effectively improved the fault tolerance and efficiency of model training\footnote{Details will be included once the paper is accepted.}.

\section{Preliminaries and Motivations}\label{sec:back}
This section introduces the preliminaries of this work and the limitations of existing studies to motivate~\name.

\subsection{Parallel Training Techniques for Large Models}

\paragraph{Hybrid Parallel Training}


HP is a prevalent method for distributed Large Model (LM) training \cite{narayanan-sc21}, combining three main parallelism techniques: data parallelism (DP), tensor model parallelism (TP), and pipeline parallelism (PP).
\textit{DP}~\cite{imagenet-cnn-cacm17,all-reduce-16,geeps-eurosys16,poseidon-atc17,tang2020survey,GossipFL} replicates a model across multiple devices to enable parallel training, which can lead to parameter redundancy. To address this issue, ZeRO-DP has been introduced~\cite{rajbhandari2020zero}, including variants like ZeRO-1, ZeRO-2, and ZeRO-3, which distribute parameters across different DP paths. Among these, ZeRO-1 introduces less communication overhead and is preferred for scaling HP.
\textit{TP}~\cite{dean-nips12,parallelize-cnn14,elastic-mlsys20,corr/abs-1909-08053} distributes the computation of large tensors among multiple devices, enabling the processing of models that exceed the memory capacity of a single device. Each device performs computations independently and communicates as needed.
\textit{PP}~\cite{pipedream-sosp19,dorylus-osdi21,deepspeed-sc20,tang2023fusionai} divides model layers into stages assigned to different devices. Each stage processes intermediate results from the previous device and forwards its output to the subsequent stage.

In practice, large-scale distributed training typically leverages vanilla DP or ZeRO-1 with TP and PP to ensure the scalability of HP. This research aligns with the common approach of integrating these parallelism techniques for effective LM training.

\paragraph{Failures in Hybrid Parallelism}
Researchers have observed a high incidence of system failures during LM training, especially as the system size expands to involve thousands of GPUs~\cite{zhang2022opt,jiang2024megascale}. Although these failures may incur on a small scale~\cite{jiang2024megascale}, they require a complete system restart from the latest checkpoint. This restarting process incurs significant recomputation overhead, particularly when the latest checkpoint was created hours before. 

There are two primary categories of failures encountered during LM training: hardware and software failures~\cite{zhang2022opt,234916,Li_2023}. Hardware function failures, often triggered by temperature fluctuations, power disruptions, and storage inconsistencies, are prevalent in LM pretraining~\cite{zhang2022opt,Li_2023}. On the other hand, software failures are common due to frequent inter-device communications and high memory demands, which put a strain on software components like MPI~\cite{Forum_1994} and PyTorch memory management~\cite{pytorch}.

\subsection{In-Memory Checkpointing for LM Training}
{In-memory checkpointing for LM training can be implemented through two primary methods: \textit{asynchronous parameter snapshotting} and \textit{in-memory checkpoint protecting}. }

\begin{figure}[t]
\centering
  \includegraphics[width=0.47\textwidth]{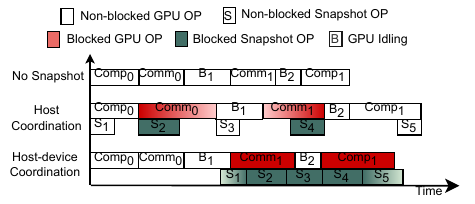} 
\caption{Resource competition between training
and snapshotting 
}
\label{fig:async}
\end{figure}

Asynchronous parameter snapshotting can lead to competition for memory bandwidth with training processes, potentially slowing down the training speed. Existing studies mainly introduce two ways to coordinate snapshotting and training to reduce interference. As shown in Figure~\ref{fig:async}, \textit{Host Coordination}~\cite{mohan_checkfreq_nodate,torchsnapshot} uses a separate process or thread instructing asynchronous snapshotting without considering device scheduling. 
Conversely, Gemini~\cite{gemini} and TorchSnapshot~\cite{jiang2024megascale} introduce \textit{Host-Device Coordination}, which leverages the asynchronous capabilities of PyTorch~\cite{pytorch} to allow computations on device (mainly for training) and memory transfers (mainly for snapshotting) to occur simultaneously.
%
In-memory checkpoint protecting is proposed by Gemini~\cite{gemini} in ZeRO-3 to utilize inter-node traffic to distribute local parameter shards onto peer nodes. Gemini's design allows training and inter-node checkpoint communications to occur concurrently.


\begin{figure}[tb]
\centering
  \includegraphics[width=0.47\textwidth]{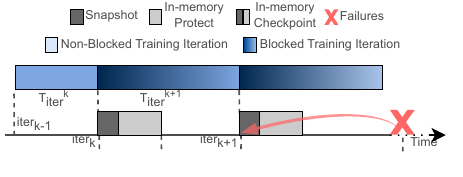} 
\caption{Asynchronous in-memory checkpoint overhead 
}
\label{fig:stall}
\end{figure}

When doing checkpointing, we care about two metrics, namely the in-memory checkpointing overhead $O_{in-mem}$ and the failure recovery time $O_{restart}$.
As shown in Figure~\ref{fig:stall}, $O_{in-mem}$ is measured as the \textit{delayed} training time during checkpointing and is calculated as $T_{iter}^{k+1}-T_{iter}^k$, where $T_{iter}$ represents the per iteration training time.
$O_{restart}$ measures how fast the training can recover from a failure and is calculated as the gap between the time of failure and the time of latest snapshot.
When asynchronous snapshotting is inefficient, accumulated $O_{in-mem}$ can be large. Consequently, the snapshotting frequency must be reduced to reduce the blocking of training. On the other hand, reducing snapshotting frequency leads to higher $O_{restart}$. 
In this paper, we aim to reduce $O_{in-mem}$ to allow frequent in-memory checkpoints and hence reduce failure recovery time (i.e., $O_{restart}$).

\subsection{Limitations and Motivations}

As shown in Figure~\ref{fig:intro:sota}, applying existing in-memory checkpointing techniques for LM training under HP introduces significant overhead.
There are mainly two reasons.


\paragraph{Limitation in Existing Snapshotting}

Existing in-memory snapshotting methods can be efficient when the checkpoint size is small or the GPU utilization is low~\cite{mohan_checkfreq_nodate,torchsnapshot}. For example, the host-device coordination is evaluated to be efficient in ZeRO-3~\cite{gemini}, which relaxes constraints for asynchronous snapshotting when local checkpoint sizes are small, hardware FLOPs utilization is low, and iteration time is long~\cite{zero}.
However, when training LM with HP, the FLOPs utilization becomes high, and each device's snapshot size becomes large~\cite{jiang2024megascale}. 
Larger checkpoint sizes require more snapshotting operations, which necessitates the device (i.e., GPU) to execute more \textit{async copy} instructions from the host (i.e., CPU) and to consume significant PCIe bandwidth.

\paragraph{Limitation in Existing Snapshot Protecting}
In-memory checkpoints are unreliable during hardware failures. 
The current in-memory protecting method in Gemini~\cite{gemini} distributes checkpoint shards onto peer nodes to improve reliability, which, on the other hand, introduces communication overhead. We found that this method can be efficient only under the ZeRO-3 DP training mode. With HP training, parameters are redundant or partially redundant across all DP group members in ZeRO-1 and vanilla DP~\cite{jiang2024megascale}, which are common settings in HP.
When the checkpoint size on each node grows, the communication size across peer nodes to distribute checkpoint shards increases dramatically.
Moreover, HP training requires intensive inter-node communication in pipeline and data parallelism contexts, which further exacerbates the competition for communication bandwidth, leading to a decline in training performance.


The above limitations motivate us to design \textit{new in-memory snapshotting and protecting optimizations that alleviate the resource competition between in-memory checkpointing and training}, allowing fast and reliable LM training under hybrid parallelism.

\section{Design Overview}

\begin{figure}[t]
\centering
  \includegraphics[width=0.49\textwidth]{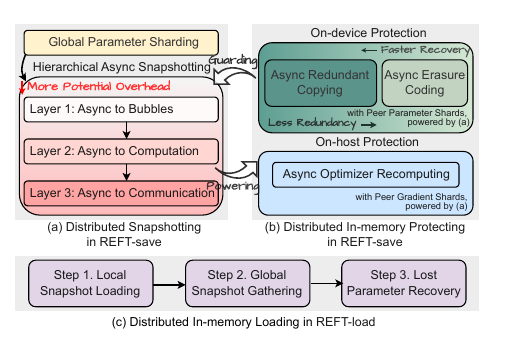} 
\caption{System Overview: \names~employs two engines to snapshot and protect parameters in host memory, generating in-memory checkpoints. 
\namel~utilizes \textit{Distributed In-memory Loading Engine} for fast failure recovery.}
\label{fig:designall}
\end{figure}

To design an efficient in-memory checkpointing system for hybrid parallel training that alleviates resource competition, we propose \name.
As shown in Figure~\ref{fig:designall}, \name~mainly has two components, including \names~for checkpoint saving and \namel~for checkpoint loading.
We address the resource competition limitations mainly in the \names~component, by addressing the following questions.

\ding{182} \textbf{How can the host and devices collaborate efficiently to avoid snapshotting overhead? }

By understanding how asynchronous snapshotting and training operations interfere with each other, we propose allocating snapshotting processes in parallel with training based on interference levels.
As shown in Figure~\ref{fig:designall}(1), after parameter sharding, distributed training devices perform \textit{Hierarchical Asynchronous Snapshotting (HAS)} when \names~is called. 
HAS includes three layers when scheduling snapshotting processes. Snapshotting is first allocated to device bubbles (i.e., device idling). If bubbles are insufficient, \name~uses layer 2, separating snapshotting processes into different computation periods to reduce resource competition. Finally, snapshotting is assigned to periods of high-speed intra-node communications that do not use PCIe. Profiling in Section~\ref{sec:design} shows this may introduce overhead, so we try to avoid this option. With \textit{HAS Coordination}, \name~fully utilizes the idled device time to reduce contention.

\ding{183} \textbf{How can we provide minimal redundancy protecting overhead while enhancing the completeness of distributed checkpoints? }

As shown in Figure~\ref{fig:intro:sota}, frequent inter-node communications for in-memory checkpoint protecting block training. To address this, the in-memory protecting in \names~introduces redundancy across peer nodes to seek optimization opportunities.
Note the HAS coordination creates redundancy at the intra-node level by capturing snapshots of parameters for which peer nodes are responsible for saving.
This method doubles the snapshot size but allows for fast recovery from hardware failures.
However, should this redundancy lead to excessive overhead, the system can switch to \textit{Asynchronous Erasure Coding (AEC)}, which convert local snapshot segments into parity data, thereby reducing the load on the PCIe. While recovery with AEC is slower due to the need for parity decoding, it effectively reduces the burden of frequent snapshot saving.

For optimizer parameters that are not redundant in sharded DP optimizations (e.g., ZeRO-1 DP), each node uses \textit{Asynchronous Optimizer Recomputing (AOR)} to capture gradient shards assigned to peer nodes for snapshotting and reconstruct the peer optimizer parameters locally. This approach ensures parameter redundancy in the collective memory of the hosts, even for parameters that are inherently not redundant.
The system allows for a dynamic combination and switching between these distributed in-memory protection methods, which can be configured either by user settings or automatic profiling at the start of training. As shown in Section~\ref{sec:evaluation}, the integration of these protection methods to enhance the reliability of in-memory checkpointing results in minimal to no overhead.

In the following section, we will detail the design of \name, focusing on leveraging available hardware resources in hybrid parallelism for efficient in-memory checkpointing. With protected in-memory checkpoints, we can let nodes communicate their local parameters with peers on node failures via Ethernet and restart the system without accessing the slow NFS. We will illustrate this process in detail in Section~\ref{sec:design}.

\section{REFT: Reliable and Efficient In-memory Fault Tolerance}\label{sec:design}
\begin{figure}[t]
\centering
  \includegraphics[width=0.45\textwidth]{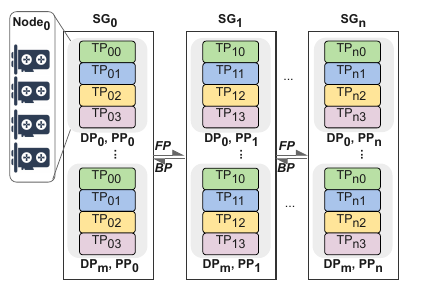} 
\caption{An $SG$ (Sharding Group) refers to a group where the parameters of assigned partitions are saved. In this example, we use TP intra-node and PP inter-node, a common 3D parallel setting~\cite{narayanan-sc21,zhang2022opt}. The model partition across TP and PP within the same DP differs. In 3D parallel pretraining, micro batches are fed into the LLM from the left side, passing through forward and backward passes. All nodes in the same PP state form an $SG$, e.g., all $PP_0$ nodes form $SG_0$.}
\label{fig:3dp}
\end{figure}
\name~ is designed to efficiently and reliably scale hybrid parallelism for fault-tolerant language model (LM) training with efficient saving and loading functions, as in Figure~\ref{fig:designall}. In hybrid parallel deployments, which include combinations of Data Parallelism (DP), Tensor Parallelism (TP), and Pipeline Parallelism (PP), \name~aims to minimize in-memory checkpointing overhead, allowing training to resume from the most recent iteration with minimal loss of progress in the event of failures.
To achieve these goals, \name~provides two important modules: \names~optimizes in-memory checkpoint saving with \textit{Distributed Snapshotting} and \textit{Distributed In-memory Protecting}; \namel~proposes in-memory checkpoint loading with \textit{Distributed In-memory Loading}.

\subsection{Distributed Snapshotting with REFT-save}
The \textit{Distributed Snapshotting Engine} of \names~aims to maximize the utilization of the distributed parallelized device-to-host communication and storage I/O of the entire system, and implement asynchronous snapshotting with minimal disruption to LLM pretraining. 

\paragraph{Global Parameter Sharding}
In hybrid parallel training, we define the standard sharding process for parallel checkpoint saving. 
As shown in Figure~\ref{fig:3dp}, in hybrid parallelism, each pipeline parallel stage is designated as a sharding group $SG$. Parameters to be divided in the $n$-th $SG$ are denoted as $W_n$. With a total of $m$ DP nodes (i.e., nodes within the same PP stage), each stage on a DP has $\frac{W_n}{m}$ parameters, reducing the data transmission load on each node by a factor of $m$ during checkpointing. In this work, for DP in hybrid parallelism, we focus on vanilla DP and ZeRO-1, which scale training efficiently and are commonly used for large-scale LM training. In ZeRO-1~\cite{rajbhandari2020zero}, the scope of sharding is limited to model parameters because optimizer parameters are initially partitioned across DP paths. Only model parameters are redundant on each DP path and need to be sharded for parallel checkpointing. \name~will check for redundancy across all devices. If found, we group these devices into a sharding group. \name~then shards parameters to save within this group.

\paragraph{Hierarchical Asynchronous Snapshotting (HAS)}


\begin{algorithm}
\caption{Hierarchical Asynchronous Snapshotting}
\begin{algorithmic}[1]
\State \textbf{Input:} 
\State $\mathcal{P}$: Pipeline stages, $\mathbf{W}$: Whole model parameters;
\State $B_{io}$: IO bandwidth;
\State $C_{FB,BP}$: Forward and backward time of one micro batch;
\State \# Scheduling snapshot based on runtime estimation:
\For{pipeline stage $i \in \mathcal{P} $  in parallel}
    \State $t_{ss} \leftarrow $ EstimateSnapshotTime($i$, $\mathbf{W}$, $B_{io}$); 
    \State $t_{bubble} \leftarrow $ EstimateBubbleTime($i$, $C_{FB,BP}$); 
    \If{$ t_{ss}  \geq t_{bubble}$}
        \State $\mathbf{W}^{i}_{bubble}, \mathbf{W}^{i}_{*} \leftarrow $ SplitParameter($\mathbf{W}^i, t_{ss}, t_{bubble}$)
    \Else
        \State $\mathbf{W}^{i}_{bubble}, \mathbf{W}^{i}_{*} \leftarrow \mathbf{W}^i, None$
    \EndIf
    \State Launch async snapshot $\mathbf{W}^{i}_{bubble}$ in bubble time;
    \State Launch async snapshot $\mathbf{W}^{i}_{*}$ in computation time.
\EndFor
\State
\State \textbf{def} \texttt{SplitParameter}($\mathbf{W}, t_{ss}, t_{bubble}$):
\State \ \ \ \ $n \gets \text{number vectorized} \ \mathbf{W}$
\State \ \ \ \ $\mathbf{W}_{bubble} \gets \mathbf{W}[: \lfloor n \times t_{bubble}/t_{ss}  \rfloor]$
\State \ \ \ \ $\mathbf{W}_{*} \gets \mathbf{W}[ \lfloor n \times t_{bubble}/t_{ss} \rfloor:]$
\State \ \ \ \ \textbf{return} $\mathbf{W}_{bubble}, \mathbf{W}_{*} $

\State \textbf{def} \texttt{EstimateSnapshotTime}(stage $p$, $\mathbf{W}$, $B_{io}$):
\State \ \ \ \ \textbf{return} size($\mathbf{W}_p$) / $B_{io}$

\State \textbf{def} \texttt{EstimateBubbleTime}(stage $p$, $C_{FB,BP}$):
\State \ \ \ \  \textbf{return} ($0.8 p + 2|\mathcal{P}| - p -2) \times C_{FB,BP}^p$

\end{algorithmic}
\label{alg:async}
\end{algorithm}

To address the asynchronous snapshotting performance discussed in Section~\ref{sec:back}, the engine strategically minimizes interference between snapshotting and training operations. After the Global Parameter Sharding process that shards the snapshotting processes across all devices for parallel snapshotting, the engine structures snapshotting into three layers, as presented in Algorithm~\ref{alg:async}. The layers determine the exact timing and sizes for each snapshotting operation:

\ding{182} Layer 1 performs snapshotting asynchronously during \textit{device bubbles} to minimize overhead. The bubble is defined as when the device is not computing or communicating. This method is especially effective in pipeline parallelism when device bubbles are evident. It achieves this by profiling each bubble's available bubbles and checkpoint parameters.

\ding{183} Layer 2 conducts snapshotting in parallel with the \textit{computation stream} if pipeline bubbles are insufficient. Due to a reduced snapshotting operation filtered to this layer, the possibility of interference between computation and snapshotting is small, thus avoiding training disruptions. 

\ding{184}  Layer 3 operates asynchronously to the \textit{communication stream}. This layer only works when training and snapshotting use separate interconnections (NVLink for training and PCIe for snapshotting) to reduce possible conflict. Note that this layer will not be used unless the previous layers are not enough for snapshotting operations.

Algorithm~\ref{alg:async} demonstrates the process of scheduling asynchronous snapshotting with the training process. For each stage $i$, it first estimates two key times: the snapshot time $t_{ss}$ and the bubble time $t_{bubble}$. The snapshot time $t_{ss}$ is calculated based on the IO bandwidth $B_{io}$ and the communication costs of the model parameters. The bubble time $t_{bubble}$ is estimated considering the computational costs for forward and backward propagation $C_{FB, BP}$ and the GPU computational FLOPs $C_{gpu}$. Note that here the bubble time estimation is specifically tailored for the 1F1B pipeline parallel structure. We also provide an online bubble profiler for other pipeline parallel configurations; we can accurately estimate the computation time through runtime profiling, which facilitates the scheduling of snapshots.

If the snapshot time $t_{ss}$ is greater than or equal to the bubble time $t_{bubble}$, the algorithm splits the parameters into two parts: $\mathbf{W}_{bubble}$ and $\mathbf{W}_{*}$. The algorithm prioritizes exploiting bubble time to snapshot $\mathbf{W}_{bubble}$. Then, the left part of the model $\mathbf{W}_{*}$ is snapshotted within the computation time. On our testbed in Section\ref{sec:method}, HAS introduces little to no overhead to training on large-scale hybrid parallel training settings.


\subsection{Distributed In-memory Protecting with REFT-save}
In-memory checkpoints broken by a single node failure can lead to incomplete global checkpoints, preventing a fast restart. Thus, in-memory checkpointing methods must generate redundancy for identical checkpoints.
As discussed in Section~\ref{sec:introduction}, communication between nodes to generate such redundancy does not apply to hybrid parallelism, leading to large training overhead.
So, \name~chooses to generate redundancy with intra-node operations.  
For different hybrid parallelism, the \textit{Distributed In-memory protecting Engine} ensures redundant protecting for in-memory checkpoints across distributed nodes, leveraging available device and host resources to optimize parameter safety without increasing training overhead.  

\begin{figure}[t]
\centering
  \includegraphics[width=0.45\textwidth]{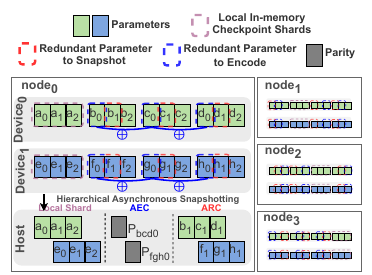}
  \caption{Two in-memory checkpoint protection methods, AEC and ARC, are used in a system with four nodes, each having two GPUs. Both AEC and ARC are designed to avoid intra-device communication, which could block HP in hybrid parallelism. In addition to the local shard that needs to be snapshotted, Node 0 uses AEC to snapshot parities, reducing the size during redundant snapshotting. ARC, on the other hand, redundantly copies the parameters of other nodes responsible for snapshotting.}
  \label{fig:hybrid}
\centering
\hfill
  \includegraphics[width=0.43\textwidth]{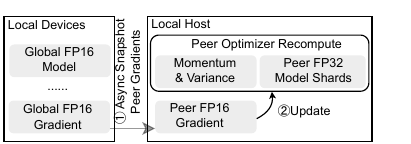}
  \caption{The process of AOR in ZeRO-1 DP.}
  \label{fig:aor}
\end{figure}


\paragraph{Asynchronous Redundant Copying (ARC)}
In ARC, when inherent redundancy exists among the parameters in a sharding group $SG$, each member not only saves its own shard to a snapshot but also saves shards from peer members in the $SG$ to generate redundancy in the global host memory. Then, in the event of a failure, parameters lost on a failed node can be restored from their redundant duplication on other nodes. 
The red boxes in Figure~\ref{fig:hybrid} give an example of ARC.
Given a PP size $n$ and DP size $m$, ARC doubles the volume of snapshots to $2\frac{W_n}{m}$ for each group member in the $SG$. This burden could introduce large overheads to other asynchronous snapshotting methods, but with HAS coordination, the overhead could be reduced up to zero.
With this method, the training process can be restored as long as there is no more than one node failure within the $SG$. 

\paragraph{Asynchronous Erasure Coding (AEC)}

    
    


To reduce the size of redundancy when it introduces snapshotting overhead, we propose to use Asynchronous Erasure Coding (AEC) as the second method for protecting the engine. 
The AEC relies on the logical computation capability of GPU devices and is embedded in device bubbles. In our micro-benchmarks, on-device erasure coding achieves up to 12$\times$ to 15$\times$ higher throughput than snapshotting on V100 and A100 GPUs.
The blue boxes in Figure~\ref{fig:hybrid} illustrate an example of AEC in the system with four nodes. Same as ARC, to prevent inter-node communication from interrupting the PP communications, AEC snapshots $a_0$, $a_1$, $a_2$ on one device redundantly protect $b_0$, $c_0$ and $d_0$ with the following encoding method: 

\textit{i. Encoding.}
In training bubbles on device 0, \name~compute the parity 
$p_{bcd0}$ of $b_0$, $c_0$ and $d_0$ by encoding the parity unit with XOR calculations as:
\begin{equation}
    p_{bcd0} = b_0 \oplus c_0 \oplus d_0
\end{equation}
The same encoding process is executed on peers in the $SG$. 

\textit{ii. Decoding.}
When a node in an SG experiences offline failure, \name~can restore the parameters on the node with XOR calculations. Assuming $node_0$ is offline, the system retrieves $a_0$ using the the subtraction decoder: 
\begin{equation}
a_0 = p_{abc0} \oplus b_0 \oplus c_0 
\end{equation}
The same procedure applies to other parameters. 
Notably, Given a PP size $n$ and DP size $m$, AEC necessitates redundant parity snapshotting that reduces the volume of redundancy to $\frac{W_n}{m(m-1)}$. The reliability of AEC is the same as ARC.

\paragraph{Asynchronous Optimizer Recomputing (AOR)}
In optimizer sharded Data Parallelism (DP), such as ZeRO-1, optimizer parameters are evenly distributed among the DP group members, losing inherent redundancy. However, model parameters and gradients remain complete on each member. To provide redundancy protecting without inter-node communication overhead or extra device computation, we introduce an on-host optimizer protecting module named Asynchronous Optimizer Recomputing (AOR). Specifically, with ZeRO-1 enabled, the system maintains replicas of optimizer parameters from peer members in the DP group. Figure~\ref{fig:aor} gives an example and AOR.
Specifically, each node in the hybrid parallelism (identified as $i$, belonging to pipeline stage $j$) is responsible for protecting peer shards of the same size as local shards. Given the total model parameters $W{\text{model}}$ and total optimizer parameters $W{\text{optimizer}}$, distributed across $n$ pipeline stages each having $m$ data parallel members, the model and optimizer parameter shards managed by each node are expressed as:
\begin{equation}
    W_{\text{model, shard}}^{(i, j)} = \frac{W_{\text{model}}}{mn}, \quad W_{\text{optimizer, shard}}^{(i, j)} = \frac{W_{\text{optimizer}}}{mn}
\end{equation} 
where $i = 1, 2, \ldots, m$ and $j = 1, 2, \ldots, n$. 
Each node computes the gradient of its model parameter shard $\nabla W_{\text{model, shard}}^{(i, j)}$ on local devices and uses the hierarchical asynchronous method to snapshot the gradient to update its corresponding optimizer parameter shard on the host:
\begin{equation}
W_{\text{optimizer, shard}}^{(i, j, t+1)} = W_{\text{optimizer, shard}}^{(i, j, t)} - \eta \nabla W_{\text{model, shard}}^{(i, j, t)}
\end{equation}
where $\eta$ is the learning rate, and $t$ indicates the time step of the update. This update uses the redundant host FLOPs and is asynchronous to training. On failures, the system retrieves optimizer parameters from host memory with redundant parameters.

\paragraph{Collaborative Redundant Protection}
If $N$ redundant protection strategies are enabled simultaneously, or the same redundant protection strategy is used $N$ times with different parameters, the protection of our redundant protection system can be enhanced by $N-1$ times without any overhead. If $N$ or fewer node failures occur in the same $SG$, the training process can be restored. We also evaluate the efficiency of collaborative protection in Section~\ref{sec:evaluation}.

\begin{figure*}[t]
\centering
  \includegraphics[width=1\textwidth]{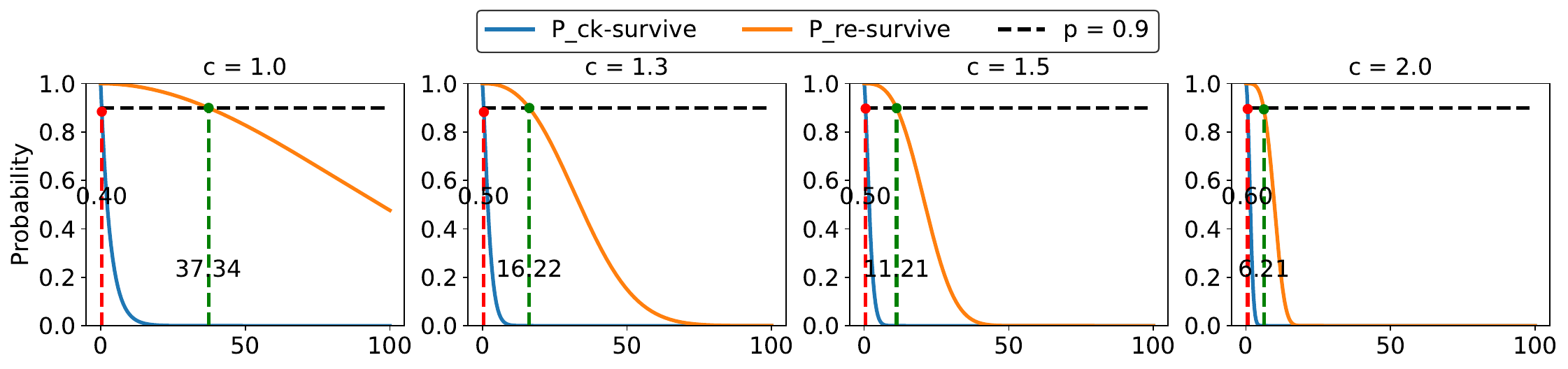} 
\caption{We compare \name's parameter survival probability with checkpointing on a 3072-device system with 6 DP paths, similar to Megatron. With a hardware failure rate of 0.0001, a software failure rate of 0.00001, and varying shape parameters (c = 1.0, 1.3, 1.5, 2.0, sampled from \cite{laosooksathit2014reliability}). \name~significantly boosts survival probability. For example, given a survival threshold of 0.9 and a parameter c = 1.3, \name~necessitates a checkpoint only once every 16.22 days thanks to the erasure coding that improves the survival rate in Equation 3. In contrast, checkpoint-based methods call for one every 0.5 days. This implies that with \name, parameters can persist safely in the volatile host memory for 16.22 days, whereas without \name, they last just around 0.5 days before becoming unsafe. Note that the numbers in this figure are based on assumptions of the failure rate. The actual failure rate could be larger and require more frequent checkpointing, based on the observation in \cite{zhang2022opt}.}
\label{fig:prob}
\end{figure*}

\subsection{Distributed In-memory Loading with REFT-load}

As shown in Figure~\ref{fig:designall}, \namel~is powered by the \textit{Distributed In-memory Loading Engine} that manages the synchronization of parameters after failures from global host memory loading with three steps. It avoids the inefficiency of external Network File System (NFS) storage read by: \ding{182} Directing each node to load its checkpoint shard from local Host memory. \ding{183}  Facilitating an all-gather operation among nodes to synchronize parameters across the network. \ding{184}  Decoding or transferring missing parameters whenever a node experiences parameter loss, ensuring that all nodes can effectively recover and synchronize their training state.
These components collaborate to ensure that \name~delivers a scalable and reliable fault-tolerant training environment in hybrid parallelism, maximizing in-memory checkpointing efficiency. However, in case the number of failed nodes exceeds the capacity of in-memory protecting, \name~allows training to restart from global checkpoints in NFS. Our analysis in Section \ref{sec:evaluation} shows that the reliability of in-memory checkpoints in REFT is adequate for LM training in most cases. The following section offers an in-depth design of how \name~works.

\subsection{Workflow}
\paragraph{Elastic Functionality}
\name~operates elastically on rendezvous function of TorchElastic\cite{pytorch-elastic}. Upon training initiation, a HEALTHY signal triggers the start of training. When it receives the SNAP signal for snapshotting to begin, REFT-save starts and host memory saves parameters asynchronously. Shards are flagged as COMPLETED when all tensors are snapshotted. On failure, an UNHEALTHY (software failure) or OFFLINE (node failure) signal is broadcast from the failure node, and training restarts elastically. If the parameters are recoverable, REFT-load will attempt to recover them during the restart.
\paragraph{Snapshot Management}
In \name, snapshots are saved on redundant host memory and flushed to tmpfs~\cite{Snyder1990tmpfsAV} as needed. Normally, it maintains at least one completed copy and one ongoing snapshot. The number of completed copies is limited by the assigned host memory to avoid Out-of-memory issues. The completed and ongoing snapshots are managed to maintain parameter consistency under the following conditions:
\ding{182} \textit{Saving:} The ongoing snapshot accepts flushed parameters from device memory. When saving is complete, the completed snapshot is replaced by the new copy of the ongoing snapshot. This cycle prevents parameter inconsistency that may harm LLM convergence. Note that \name~saves in-memory checkpoints to NFS at lower frequency to guard against large-scale hardware failures.
\ding{183} \textit{Loading:} When pretraining processes suddenly fail or stop, the REFT-save persists the latest completed snapshots and their redundancy to tmpfs. If the failure is at the hardware level, causing nodes to restart or shut down, parameters are reconstructed from redundant backups after nodes all-gathering. If reconstruction fails, training restarts from an existing checkpoint in NFS.

\section{System Reliability Analysis}

\begin{assumption}\label{ass1}

In multi-node GPU systems, we assume all nodes' failure probabilities are independent. These systems' Time-to-Failure (TTF)—the time from system operation onset to the first failure—follows a Weibull distribution. The Weibull distribution is a statistical model widely used to describe the lifetime and failure patterns of systems, commonly applied in failure modelling scenarios~\cite{10.1177/1094342012464506}\cite{Schroeder_Gibson_2006}.  
\end{assumption}

\textbf{Time-to-Failure (TTF)}. TTF refers to the duration a system or component operates reliably. We define the system’s failure rate as $\lambda_{fail}$. Based on our assumption and the fact that the system has been running for time $t$, the cumulative probability of survival is given by
$ P = e^{-\lambda_{fail}t^c}$,\noindent where $c$ is the shape parameter of the Weibull distribution, influencing the distribution's shape and, consequently, the failure probability pattern. Suppose there are $k/n$ $SG$s in a system of $k$ nodes. We refer to the parameters captured in the latest snapshot as the current parameters.

\textbf{Survival Probability with \name}. \name~provides maximum in-memory saving efficiency and near to zero overhead to training, thus allowing iteration-level saving. Then, the survival probability of single-layer redundancy protecting with AOC or AEC in \name, denoted as $P_{\text{re-survive}}$ ($re$ stands for \name), indicates the likelihood that \name~successfully protects in-memory checkpoints from being lost, assuming all submodules function correctly and at most one node fails in each $SG$. 
This probability is independent of training processes. 
The overall probability of the parameters surviving a multi-node system failure and efficiently restarting from in-memory checkpoints with \name~is expressed as:

\begin{equation}
P_{\text{re-survive}} = (P_{s}^n + n(1-P_{s})P_{s}^{(n-1)})^{\frac{k}{n}}P_{re}^{k}
\end{equation}

Here, $P_s$ is the cumulative probability of a single node surviving hardware failures, and $P_{re}$ is the probability of a single node's SMP program failing. $1-P_s$ represents the probability of no node failure at time $t$.

\textbf{Survival Probability without \name}. Without using \name, pretraining cannot survive on node failures and have to experience NFS checkpoint loading. The probability of surviving without \name~is predicated on all nodes being healthy:

\begin{equation}
P_{\text{ck-survive}} = P_s^k P_{tr}^k
\end{equation}

where $P_{tr}$ represents the cumulative probability of a single node surviving software failures. The probabilities $P_{\text{ck-survive}}$ and $P_{\text{re-survive}}$ over time $t$ are depicted in Figure \ref{fig:prob}. We can conclusively state that \name~significantly enhances survival probability by safeguarding parameters, and with \name, the frequency of NFS checkpointing can be greatly reduced.

\section{Implementation}\label{sec:implement}
\name~supports hybrid parallelism with scalable efficiencies, combining DP, TP, and PP. It is implemented on DeepSpeed v0.12.4 and PyTorch v2.2.1, with over 4000 lines of Python codes, integrating in-memory checkpoint snapshotting, protecting, and loading. Its core function is pluggable for other deep learning training frameworks built on PyTorch.
For asynchronous snapshotting optimizations, \name~employs multi-threading on the host side for asynchronous coordination. It enables pin memory on host buffers for faster device-to-host data transfer. On the device side, it uses multiple CUDA streams and ROCm processes for coordination. For checkpoint loading and protecting, communication between GPU devices is handled via NCCL in CUDA and RCCL in ROCm. Erasure coding computations are implemented by encoding FP32 or FP16 parameters with bitwise XOR. The implementation strictly follows \name's fine-grained design, minimizing checkpoint saving and loading overhead during training.

\section{Evaluations} \label{sec:evaluation}
\subsection{Evaluation Setups} \label{sec:method}
\paragraph{Testbeds}
We evaluate \name~under various hybrid parallelism configurations. \name~is designed and optimized on NVIDIA GPUs, and evaluated at scale on the NVIDIA Cluster. We also present the generalization capability of \name~on Frontier AMD GPU cluster. 
The hardware configurations are detailed in Table~\ref{tab:hard}. Our evaluation uses 512 GPUs from each cluster. On the NVIDIA Cluster, each V100-SXM2-16GB is connected to three GPUs via 300GB/s P2P NVLink bandwidth, and the nodes are connected via Ethernet with a 100Gbps bandwidth interlink and to NFS via 10Gbps networks. On the Frontier Cluster, each MI250x device includes 2 GPUs. Each intra-node MI250X is connected to seven peers with 50-100GB/s P2P Infinity Fabric bandwidth, and the Ethernet uses 200Gbps NICs.


    
    

\begin{table}[htbp]
  \centering
  \small
  \caption{Hardware Specifications of V100-SXM2-16GB and MI250X-128GB GPU Clusters. The Bandwidth(BW) D to D is the bandwidth of the p2p communication. The peak bandwidth of MI250X ranges from 50-100 GB/s based on the number of Infinity Fabric connections between individual GPUs. All bandwidth are for unidirectional data.}

\begin{tabular}{|c|c|c|c|c|}
\hline
Device       & \begin{tabular}[c]{@{}c@{}}Device  \\  Num/Node\end{tabular} & Host Type                                                                             & \begin{tabular}[c]{@{}c@{}}BW\\ D to D\end{tabular} & \begin{tabular}[c]{@{}c@{}}BW \\ H to D\end{tabular} \\ \hline \hline
V100 & 4                                                           & \begin{tabular}[c]{@{}c@{}}Intel(R) \\ Xeon(R) \\ Gold 6148 \\ 2.40 GHz\end{tabular} & 150 GB/s                                                    & 16 GB/s                                                      \\ \hline
MI250X    & 4                                                         & \begin{tabular}[c]{@{}c@{}}Optimized \\ 3rd Gen \\ EPYC 64C\\ 2 GHz\end{tabular}     & 50-100 GB/s                                              & 72 GB/s                                                      \\ \hline
\bottomrule
\end{tabular}
\label{tab:hard}
\end{table}

\paragraph{Models and Datasets}
We evaluate the performance of \name~by pretraining the popular Llama-2~\cite{touvron2023llama} models across various scales, including Llama2-1.3B, 2.7B, 6.7B, and 34B. Details of the model parameters are listed in Table~\ref{tab:modelconfig}.
These configurations provide various workloads on up to 512 GPUs under different hybrid parallelism. We use the Wikipedia dataset~\cite{bert} for all evaluations as the input data into models.  

\begin{table}[htbp]
  \centering
  \caption{Model Sepcifications}
\begin{tabular}{|c|c|c|c|c|}
\hline
Llama-2                 & 1.3B & 2.7B & 6.7B & 34B  \\ \hline \hline
num\_layers      & 24   & 32   & 32   & 48   \\
hidden\_size     & 2048 & 2560 & 4096 & 8192 \\
num\_attn\_heads & 16   & 32   & 32   & 64   \\ \hline
\end{tabular}
\label{tab:modelconfig}
\end{table}


\paragraph{Baseline}
For asynchronous snapshotting, we use two baselines in our evaluations. These methods are generally used in the industry to reduce checkpointing overhead:
\begin{itemize}
    \item Host Coordination~\cite{mohan_checkfreq_nodate,torchsnapshot}.
    This method uses host coordination with a separate process or thread instructing asynchronous snapshotting without considering device scheduling. 
    \item Host-device Coordination~\cite{jiang2024megascale,gemini}. In this method, under the host's instruction with threading, the device schedules snapshotting and training using the PyTorch~\cite{pytorch} asynchronous ability. Specifically, this implementation enables pin memory on the host memory buffer and sets non-blocking=True for each asynchronous device-to-host copy. 
\end{itemize}

For in-memory checkpoint protection evaluations, we use (1) inter-node checkpoint communication~\cite{gemini} and (2) no snapshotting or protection as the baseline. As mentioned in Section~\ref{sec:introduction}, inter-node checkpoint communication blocks hybrid parallel training in our evaluation. We specifically compare the vanilla training performance to determine whether \name~can provide near-zero in-memory checkpoint protection overhead in large-scale LM training deployments.

\paragraph{Evaluation Metrics}
We assess \name~using three average metrics: (1) Iteration time in seconds; (2) Samples per seconds; (3) Tera Floating Point Operations per Second (TFLOPs). For different evaluations, we report selected metrics in the Figure for clarification. We use relative improvement in the metrics in these reports

Given that \name~is designed for lossless fault tolerance of synchronous parallel training, this paper does not include model convergence results. For all Llama-2 models, we use the Adam \cite{adam} optimization algorithm, which introduces triple extra parameters to save. In each experiment, we choose a batch size to maximize GPU memory use in the training system. During testing, each experiment is executed for 100 iterations, excluding a preliminary 10-iteration warm-up.

\subsection{Efficiency of Hierarchical Asynchronous Snapshotting (HAS)}
\begin{figure}[t]
  \centering
  \includegraphics[width=0.48\textwidth]{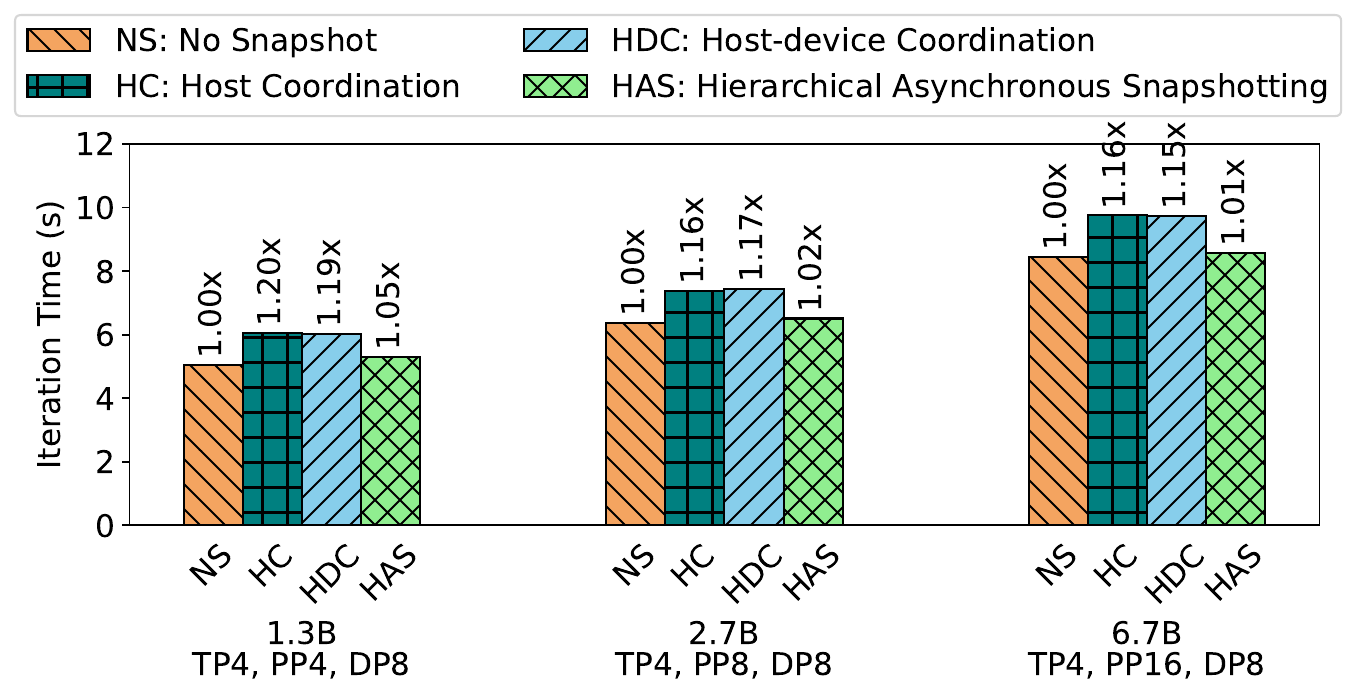}
  \caption{Iteration time comparison in hybrid parallel training on up to 512 V100 GPUs}
  \label{fig:eva1}
\end{figure}

\begin{figure}[htbp]
  \centering
  \includegraphics[width=0.48\textwidth]{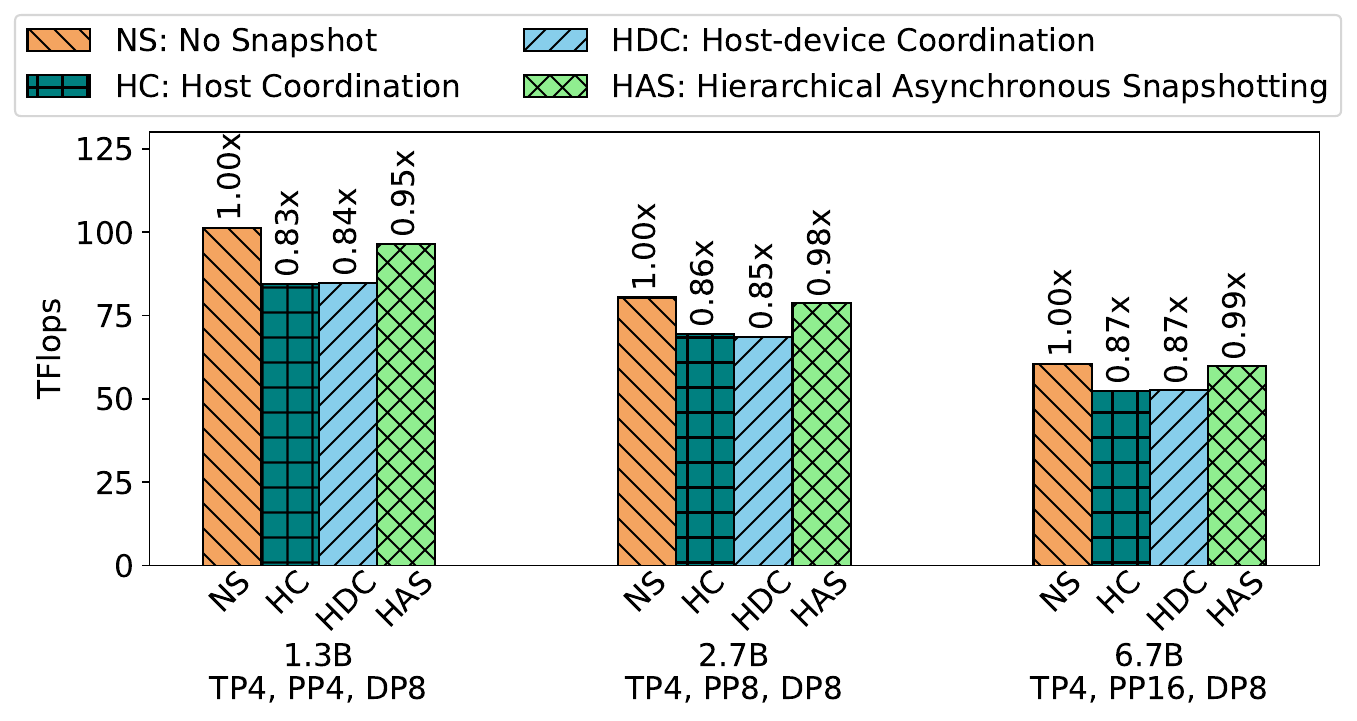}
  \caption{FLOPs comparison in hybrid parallel training on up to 512 V100 GPUs}
  \label{fig:eva2}
\end{figure}

We evaluate the efficiency of HAS by pretraining Llama-2 models on up to 512 V100 GPUs. As we can see in Figure~\ref{fig:eva1} and Figure~\ref{fig:eva2}, the iteration time and TFLOPs of HAS is over 17\% better than existing asynchronous snapshotting methods. The overhead is near zero. thus greatly improve the possible frequency of in-memory checkpointing. In the weak scaling experiment, in Table~\ref{tab:perf}, HAS also introduce less than 3\% overhead.

\begin{table}[b]
    \centering
    \begin{tabular}{|c|>{\centering\arraybackslash}p{2cm}|>{\centering\arraybackslash}p{2cm}|}
        \hline
        Hybrid Parallelism & No Snapshot & HAS \\
        \hline
        \hline
        DP1, PP16, TP4 & 7.69 & 7.63 \\
        DP2, PP16, TP4 & 15.15& 15.17 \\
        DP4, PP16, TP4 & 30.07 & 29.67 \\
        \hline
    \end{tabular}
    \caption{Performance comparison between different models (samples/s)}
    \label{tab:perf}
\end{table}


\subsection{Efficiency of In-memory Checkpoint Protecting }
In Figure~\ref{fig:protect}, we evaluate the performance of in-memory checkpoint protecting methods including ARC, AEC, and the combination of ARC and AEC. These evaluations are tested when pretraining Llama-2-34B on 512 V100 GPUs. Only 1\% overhead will be introduced if we use ARC in the experiment. The combination of ARC and AEC introduce also limited overhead to training.

\subsection{Efficiency of REFT-Load}
\begin{figure}[tb]
    \centering
    \begin{minipage}{0.23\textwidth}
        \centering
        \includegraphics[width=\textwidth]{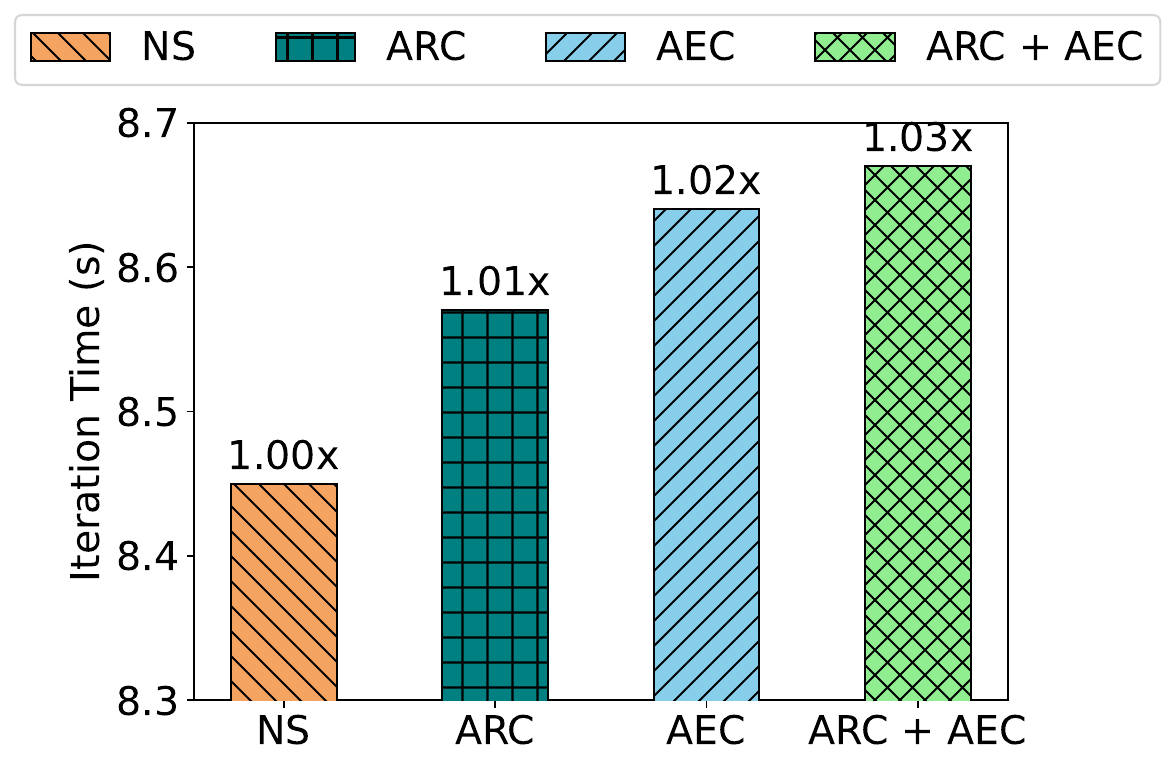}
        \caption{In-memory checkpoint protecting time of \name~when pertaining Llama-2-6.7B on 512 V100 GPUs}
        \label{fig:protect}
    \end{minipage}
    \begin{minipage}{0.215\textwidth}
        \centering
        \includegraphics[width=\textwidth]{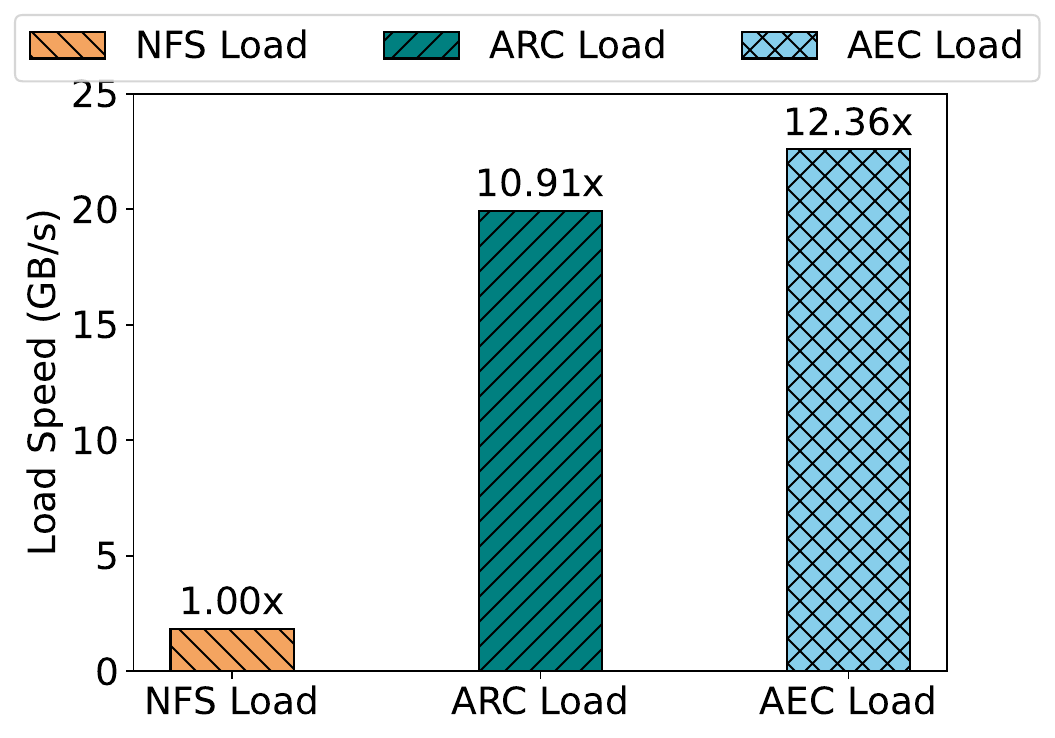}
        \caption{In-memory loading speed of \namel~when pretraining Llama-2-1.3B on 128 V100 GPUs}
        \label{fig:load}
    \end{minipage}
\end{figure}

\namel~enables hybrid parallel training to fast recover from in-memory checkpoint communications rather than NFS. As shown in Figure~\ref{fig:load}, while pretraining Llama-1.3B with TP4, PP4, DP8, we evaluate the loading performance of \namel~with different in-memory redundant protecting methods enabled.
For NFS loading, the checkpoint loading from NFS to devices took 9.27s. However, loading with ARC and AEC will take only 0.84s and 0.75s, up to 12.36$\times$ faster than NFS loading. This result validates the efficiency of in-memory checkpoint loading powered by \namel. As far as we know, \name~is the first checkpointing system that allows in-memory checkpoint loading for hybrid parallel training.

\subsection{Case Study: Generalization of REFT Optimizations on Frontier}
In the first experiment, we train a Llama-34B model with hybrid parallelism of DP 16, PP 4, and TP 8. We use ARC method to protect the model. In this experiment, the iteration time of ARC is 6.32s on average. In contrast, the average iteration time of the non-checkpointing group is 6.26s, ARC shows a slight overhead of 1\%.

In the second experiment, we train a Llama-34B model with DP 16, PP 4, TP 8 and ZeRO-1 activated. With ZeRO-1 enabled, we use AOR method to offer protection. In this experiment, the iteration time of AOR is 6.63s on average. And the iteration time of training without checkpointing is 6.59s on average. AOR method shows a slight overhead of 0.6\%.





\begin{figure}[tb]
    \centering
    \begin{subfigure}{0.23\textwidth}
        \centering
        \includegraphics[width=\textwidth]{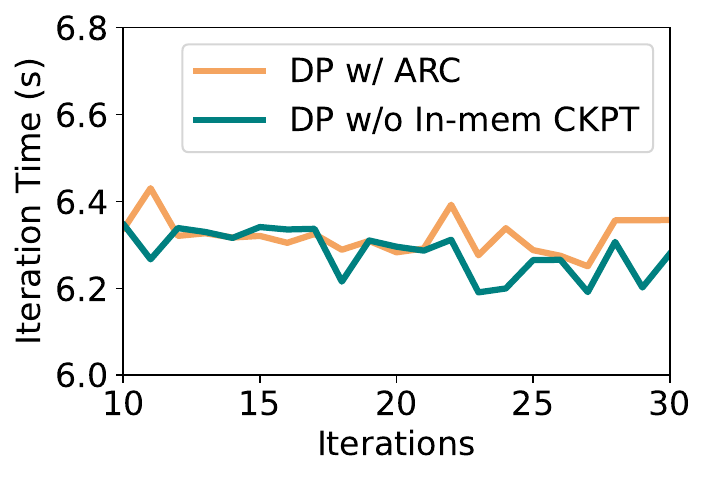}
        \caption{Iteration time with ARC v.s. without in-memory checkpointing}
        \label{fig:arc_vs_wo}
    \end{subfigure} 
    \begin{subfigure}{0.23\textwidth}
        \centering
        \includegraphics[width=\textwidth]{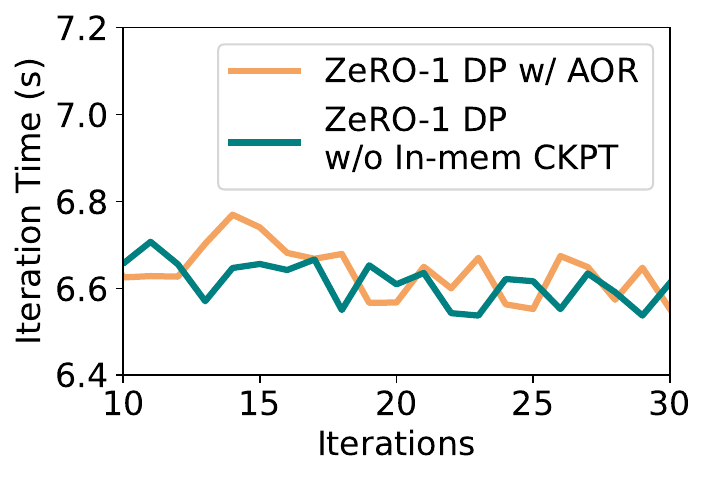}
        \caption{Iteration time with AOR v.s. without in-memory checkpointing}
        \label{fig:zero_aor_vs_wo}
    \end{subfigure}
    \caption{Both ARC and AOR methods introduce zero overhead in training and lead to an increase in training speed occasionally.
    }
    \label{fig:itertime}
\end{figure}

\subsection{Discussions}

\name~exhibit extraordinary performance in efficiency and reliability. With swift snapshotting, it can achieve optimal saving intervals and minimum fault-tolerance overhead. It is well worth mentioning that the reduced recomputation overhead on restarting also is as important as the saving efficiency, which is neglected by some papers. Still, there are some limitations of the work.

\paragraph{Insight 1: Reasonable Resource Utilization}
\name~utilizes at most 3$\times$ the storage of the optimizer and model parameters in the CPU memory, allocated for 1) the snapshotting buffer. As an illustration, in our testbed with Llama-2.7B  pretraining, the peak CPU memory usage is only 30.45GB, encompassing the data loader cache. Incorporating parameter sharding effectively mitigates the CPU memory footprint on individual nodes.

\paragraph{Insight 2: Surprising Training Acceleration}
We observe surprisingly decreased iteration time when we improve the CPU utilization by asynchronously snapshotting or conducting optimizer parameter reconstruction. This interesting observation is introduced by faster CPU operations that bottleneck training, e.g., \texttt{aten::ones} and \texttt{aten::tril} for input data loading in the first GPT layer.

\paragraph{Limitations}
Due to restricted GPU resources, we were unable to perform larger-scale pretraining. Despite this, we have proved the efficiency and reliability of \name with deployment-level GPU resources, i.e., up to 512 GPUs in Frontier.
Also, the low-level interference of asynchronous fault tolerance over training could be further mitigated by low-level programming. Since hybrid parallel training requires intensive intra-node and inter-node GPU communications, the interference of which could be significantly reduced on such training systems that leverage optimized GPU-to-GPU communication topology.

\section{Related Work}
\label{sec:related-work}

\subsection{Synchronous and Asynchronous Pipeline Parallelism}
During pretraining, pipeline parallelism may update parameters either synchronously or asynchronously. While asynchronous pipeline parallelism diminishes the bubble size in the pipeline, it does so at the expense of accuracy. For instance, PipeDream~\cite{pipedream-sosp19} updates parameters using gradients from various iterations within the pipeline. In contrast, synchronous pipeline parallelisms, as seen in Megatron~\cite{narayanan-sc21} and OPT~\cite{zhang2022opt}, perform synchronous parameter updates across DP paths. \name~facilitates consistent parameter snapshotting in synchronous pipeline parallelism during both forward and backward passes within the same iteration.

\subsection{Directions of Fault Tolerance}
There are many works using hierarchical or asynchronous methods to accelerate checkpointing. \cite{lamport, checkpointing,lamport,scr}. 

Also, previous work on recommendation model training \cite{https://doi.org/10.48550/arxiv.2104.01981}
explores the possibility of snapshot-based fault tolerance with promising results.
Also, researchers are working on optimizing the distributed checkpointing performance \cite{adam, pipedream, lazytable, scar, deepfreeze} One orthogonal direction to \name~is lossy checkpoint \cite{https://doi.org/10.48550/arxiv.1810.07354,maeng_cpr_2020}
\cite{https://doi.org/10.48550/arxiv.1810.07354} formulate the faulty with a concept of perturbation. The partial recovery is based on the fact that a part of the parameters is located in one server. Prioritized checkpoints save the parameters that have changed the most since they were previously saved. 

\section{Conclusion}

Training large models (LMs) under hybrid parallelism (HP) requires a large scale of computing resources, which, on the other hand, signifies the handling of the fault tolerance problem. 
This paper introduces \name, the first in-memory checkpointing framework for HP-based LM training, to the best of our knowledge, that experiences low or even zero checkpointing overhead.
\name~incorporates a number of system optimizations to reduce the resource competition between asynchronous checkpoint snapshotting and protecting with model training.
Thorough evaluations reveal that \name~greatly reduces parameter snapshotting and protecting overheads and allows up to iteration-level snapshotting. On failures, it efficiently rebuilds parameters from redundant parameters or parities on healthy nodes, minimizing GPU work loss. REFT has been integrated into an industrial LM training platform, which has effectively improved the fault tolerance and efficiency of model training.

\clearpage

\bibliographystyle{plain}
\bibliography{2_references}

\end{document}